\documentclass[aps,prd,preprint,superscriptaddress,showpacs,nofootinbib]{revtex4-1}
\usepackage{graphicx}
\usepackage{amsfonts}
\usepackage{mathrsfs}
\usepackage{epsfig}
\usepackage{slashed}
\usepackage{dcolumn}%
\usepackage{multirow}

\usepackage{ulem}
\usepackage{color}
\usepackage[justification=raggedright]{caption}
\usepackage{subcaption}
\definecolor{My_red}        {cmyk}{0.00,1.00,1.00,0.20}

\graphicspath{{figs/}}

\newcommand{\bmat}{\left(\begin{array}}
\newcommand{\emat}{\end{array}\right)}
\newcommand{\beq}{\begin{equation}}
\newcommand{\eeq}{\end{equation}}

\def\bwt{\begin{widetext}}
\def\ewt{\end{widetext}}
\def\be{\begin{equation}}
\def\ee{\end{equation}}
\def\bea{\begin{eqnarray}}
\def\eea{\end{eqnarray}}
\def\bean{\begin{eqnarray*}}
\def\eean{\end{eqnarray*}}
\def\bary{\begin{array}}
\def\eary{\end{array}}
\def\bit{\begin{itemize}}
\def\eit{\end{itemize}}

\def\su5u1{SU(5) \times U(1)}
\def\fsu5u1{SU(5) \times U(1)'}
\def\so10{SO(10)}
\def\sq20{SO(10) \times SO(10)}

\def\bwt{\begin{widetext}}
\def\ewt{\end{widetext}}
\def\be{\begin{equation}}
\def\ee{\end{equation}}
\def\bea{\begin{eqnarray}}
\def\eea{\end{eqnarray}}
\def\bean{\begin{eqnarray*}}
\def\eean{\end{eqnarray*}}
\def\bary{\begin{array}}
\def\eary{\end{array}}
\def\bit{\begin{itemize}}
\def\eit{\end{itemize}}

\def\su5u1{SU(5) \times U(1)}
\def\fsu5u1{SU(5) \times U(1)'}
\def\so10{SO(10)}
\def\sq20{SO(10) \times SO(10)}

\usepackage[centertags]{amsmath}
\usepackage{amssymb}

\begin{document}

\title{Isospin-Violating Dark Matter in the $U(1)'$ Model Inspired by $E_6$}

\author{Tianjun Li }
\email{tli@itp.ac.cn}

\affiliation{
	Institute of Theoretical Physics, Jiangxi Normal University,
Nanchang 330022, P.\ R.\ China
}

\affiliation{CAS Key Laboratory of Theoretical Physics and
	Kavli Institute for Theoretical Physics China (KITPC),
	Institute of Theoretical Physics, Chinese Academy of Sciences,
	Beijing 100190, P. R. China}

\affiliation{
	School of Physical Sciences, University of Chinese Academy
of Sciences, No.~19A Yuquan Road, Beijing 100049, P.\ R.\ China
}

\author{Qian-Fei Xiang}
\email{xiangqf@pku.edu.cn}

\affiliation{Center for High-Energy
Physics, Peking University, Beijing, 100871, P. R. China}

\author{Qi-Shu Yan}
\email{yanqishu@ucas.ac.cn}
\affiliation{
	School of Physical Sciences, University of Chinese Academy
of Sciences, No.~19A Yuquan Road, Beijing 100049, P.\ R.\ China
}

\author{Xianhui Zhang}
\email{zhangxianhui16@mails.ucas.ac.cn}
\affiliation{
	School of Physical Sciences, University of Chinese Academy
of Sciences, No.~19A Yuquan Road, Beijing 100049, P.\ R.\ China
}

\author{Han Zhou}
\email{zhouhan@itp.ac.cn}

\affiliation{CAS Key Laboratory of Theoretical Physics and
	Kavli Institute for Theoretical Physics China (KITPC),
	Institute of Theoretical Physics, Chinese Academy of Sciences,
	Beijing 100190, P. R. China}

\affiliation{
	School of Physical Sciences, University of Chinese Academy
of Sciences, No.~19A Yuquan Road, Beijing 100049, P.\ R.\ China
}

\date{\today}

\begin{abstract}

We propose a $U(1)'$ model inspired by $E_6$ which has an isospin-violation dark matter.
With extra two pairs of vector-like quarks, we can assign the proper $U(1)'$ charges for
the first two-generation quark doublets, and explain why the first two-generation quarks
are lighter than the third generation.
By choosing a proper linear combination of two extra $U(1)$ gauge symmetries in $E_6$, 
it is natural to realize the ratio $f_n/f_p=-0.7$ so as to maximally relax the constraints 
from the Xenon based  direct detection experiments. 
We study the sensitivities of the dark matter direct and indirect detection experiments, 
and identify the parameter spaces that can give the observed relic density.
We also study the sensitivities of the future colliders with center mass energy 
 $\sqrt{s}$= 33/50/100 TeV, and compare  the  different detection methods.
We show that in some parameter spaces the future colliders can give much stronger limits.

\end{abstract}
%11.10.Kk Field theories in dimensions other than four dimension
%11.25.Mj, Higher-dimensional gravity and other theories of gravity
%11.25.-w,Strings and branes
%12.60.Jv,Supersymmetric models 
% 12.60.-i Models beyond the standard model 
% 95.35.+d   Dark matter
%12.10.-g, unified field theories

%\preprint{}

\maketitle

\section{Introduction}

The observations of astrophysics and cosmology reveal that the main component of matter in the Universe is Dark Matter (DM). 
However, till now, all evidence for DM is through its gravitational effects, and the nature of DM particles remains a mystery.
Determining the fundamental nature of the dark matter particle is one of the most important problems in particle and astro-particle physics.
Great efforts have been taken to identify dark matter, including direct detection, indirect detection, and collider searches, 
while the answer is still unclear.

%Numerous observations confirm that our Galaxy is hosted in a DM halo, so 
DM would be detectable through their elastic scattering with nuclei in terrestrial particle detectors.
The most remarkable DM signals is the one claimed by the DAMA  Collaboration (including DAMA/NaI and DAMA/LIBRA experiment)~\cite{Bernabei:2000qi,Bernabei:2008yi,Bernabei:2010mq,Bernabei:2013xsa,Bernabei:2018yyw}, which uses a NaI-based scintillation detector.
With data collected over 14 annual cycles, the statistical significance of DAMA/LIBRA-phase2  has reached  $12.9 \sigma$~\cite{Bernabei:2018yyw}.
The CoGeNT experiment, using Germanium as a target,  also found an irreducible excess~\cite{Aalseth:2010vx} and annual modulation~\cite{Aalseth:2011wp}.
 The low energy excesses in the $\mathrm{CaWO}_4$ based experiment  CRESST-II have been reported as well~\cite{Angloher:2011uu}.
 However, these observations are challenged by the null results of the other experiments,  such as PandaX-II (2017)~\cite{Cui:2017nnn}, LUX (2017)~\cite{Akerib:2016vxi}, and XENON1T (2018)~\cite{Aprile:2018dbl}.

The Isospin-Violating Dark Matter (IVDM), in which DM couples differently to protons and neutrons,  has been proposed to reconcile the tensions among the different direct detection experimental results~\cite{ Feng:2011vu}.
Recently, the COSINE-100 experiment, that also uses the same NaI crystal as target, observes no signal excess in the first 59.5 days of data~\cite{Adhikari:2018ljm}.
This observation makes it difficult to explain all the direct detection observations, especially the observations of DAMA.
In some particular models, such as the proton-philic spin-dependent inelastic Dark Matter (pSIDM), one could still explain DAMA modulation amplitude consistent with the constraints from other experiments~\cite{Kang:2018zld}.
Here we only focus on the concept of how to realize the isospin violation in a UV complete model, instead of trying to explain all experimental observations.

Nowadays the most stringent constraint on the DM-nucleus scattering cross sections is from the Xenon based experiments~\cite{Cui:2017nnn,Akerib:2016vxi, Aprile:2018dbl}.
In this work, we will maximally relax  these constraints by naturally realizing~\cite{Feng:2011vu}
\begin{eqnarray}
\frac{f_n}{f_p}\simeq-0.7~.~\,
\end{eqnarray}

%models
Several IVDM models have been proposed in recent years. 
For scalar dark matter, Ref. ~\cite{Hamaguchi:2014pja} proposed a model with colored mediators and Ref. ~\cite{Drozd:2015gda} considered a two-Higgs doublet model. 
For Dirac dark matter, an  effective $Z'$ model was proposed in Ref.~\cite{Frandsen:2011cg}, a double portal scenario was considered in Ref.~\cite{Belanger:2013tla}, and a string-theory inspired UV model was studied in~\cite{Martin-Lozano:2015vva}.
Within the framework of supersymmetry, different realizations were examined~\cite{Kang:2010mh,Gao:2011ka,Crivellin:2015bva} . 
In this work, we propose a $U(1)'$ Model with $E_6$ origin. 
$E_6$  is of particular interesting in the sense that it is anomaly free,  
and its fundamental representation is chiral representation. 
In particular, the $U(1)'$ gauge anomaly cancellations in our models are inspired from $E_6$.
%%%

It is well-known that in the $U(1)'$ models with $E_6$ origin, the vector coupling of the up-type quarks 
to the $Z'$ boson should be zero while their axial coupling may have non-zero value. Thus, 
one cannot realize the isospin-violation with $f_n/f_p \simeq-0.7$ in the $U(1)'$ model from $E_6$.
To solve this problem, we introduce extra two pairs of vector-like quarks, and assign proper $U(1)'$ charges for
the first two-generation quark doublets. Interestingly, we can explain why the first two-generation quarks
are lighter than the third generation as well.
Considering a proper linear combination of two extra $U(1)$ gauge symmetries in $E_6$, 
we naturally realize $f_n/f_p=-0.7$ in the $E_6$ inspired $U(1)'$ model.
We  consider the constraints from dark matter direct and indirect detection experiments,  and find that there are parameter spaces in our model which can give the correct DM relic density.
Furthermore, we compare the sensitivities of the DM direct/indirect detection experiments and
the future colliders with center mass energy $\sqrt{s}$= 33/50/100 TeV. 
It is shown that in some parameter spaces the future colliders can provide much stronger limits.

The layout of this paper is as follows.
In Sec.~\ref{sec:E6}, we describe  the $E_6$-inspired $U(1)'$ model.
In Sec.~\ref{sec:DD}, we present constraints of dark matter direct detection experiments considering isospin violation effects.
Sec.~\ref{sec:collider} give the expected sensitivity of future proton-proton colliders on our model.
Finally, we conclude in Sec.~\ref{sec:summary}.

\section{ $E_6$ Inspired $U(1)'$ Model with Isospin-Violating Dark Matter }
 \label{sec:E6}

We propose the $U(1)'$ model with IVDM, which is 
a special subgroup of the $E_6$ Grand Unified Theory (GUT)~\cite{Gursey:1975ki,Achiman:1978vg,Shafi:1978gg,Ramond:1979py, Langacker:2008yv, 
Erler:2000wu, Langacker:1998tc, Erler:2002pr, Kang:2004pp, Kang:2004ix, Kang:2009rd}.  
Its fundamental representation decomposes under $SO(10)$ as

$$\bf 27 = 16  +10  +1~.$$

The representation {\bf 16} contains the $15$ SM fermions, as well as a right-handed neutrino. It 
decomposes under $SU(5)$ as 

$$\bf 16  =  10  + \bar{5} +1~.$$

The {\bf 10} representation under $SU(5)$ decomposes as

$$\bf 10 = 5  +\bar{5}~. $$

The {\bf 5} contains a color triplet and a $SU(2)_L$ doublet, whereas $\bf \bar{5}$ contains a color anti-triplet and 
another $SU(2)$ doublet, and the $\bf 1$ is a SM singlet. The gauge boson is contained in 
the adjoint $\bf 78$ representation of $E_6$.
The particle content of the $\bf 27$ representation, which contains the SM fermions as well as   
extra fermions, are shown in the first two columns of Table \ref{E6charge}. 
The SM has three generations of fermion, so  we use three such $\bf 27$. 

The $E_6$ gauge symmetry can be broken as 
follows \cite{Slansky:1981yr,Hewett:1988xc}
\begin{eqnarray}
E_6 \to\ SO(10) \times \ U(1)_{\psi} \to\ SU(5) \times\ U(1)_{\chi} \times\
U(1)_{\psi}~.~\,
\end{eqnarray}
The $U(1)_{\psi}$ and $U(1)_{\chi}$ charges for the $E_6$ fundamental ${\bf 27}$ representation 
are also  given in Table \ref{E6charge}. 

The $U(1)'$ attracting us is one linear combination of
the $U(1)_{\chi}$ and $U(1)_{\psi}$
\begin{eqnarray}
Q^{\prime} &=& \cos\theta \ Q_{\chi} + \sin\theta \ Q_{\psi}~.~\,
\label{E6MIX}
\end{eqnarray}

The other $U(1)$ gauge symmetry from the orthogonal linear combination  as well as the $SU(5)$ is  
broken at a high scale. 
This allows us to have a large doublet-triplet splitting scale, which prevents rapid 
proton decay if the $E_6$ Yukawa relations were enforced. 
This will need either two pairs of (${\bf 27}$, ${\bf {\overline{27}}}$) or one pair of (${\bf 27}$, ${\bf {\overline{27}}}$), ${\bf 78}$,
in addition  to one pair of (${\bf 351'}$, ${\bf \overline{351'}}$) dimensional Higgs representations
(Detailed studies of $E_6$ theories with broken Yukawa relations can be 
found in~\cite{King:2005jy,Babu:2015psa}.) 
For our model, the unbroken symmetry at the TeV scale 
is $SU(3)_C \times SU(2)_L \times U(1)_Y \times U(1)'$.

\begin{table}[t]
\begin{center}
\begin{tabular}{|c| c| c| c| c|}
\hline $SO(10)$ & $SU(5)$ & $2 \sqrt{10} Q_{\chi}$ & $2 \sqrt{6}
Q_{\psi}$ & $4 \sqrt{181} Q'$ \\
\hline
\multirow{3}{*}{ \bf 16}   &   ${\bf10}~ (Q_i, U_i^c, E_i^c )$ & --1 & 1  & $-9$ \\
            &   ${\bf \bar 5}~ ( D_i^c, L_i)$  & 3  & 1  & 25       \\
            &   ${\bf 1} ~(N_i^c/T)$             & --5 & 1  & $-43$         \\
\hline
\multirow{2}{*}{  \bf  10}   &   ${\bf 5}~(XD_i,XL_i^c/H_u)$    & 2  & --2 & $18$         \\
                                       &   ${\bf \bar 5} ~(XD_i^c, XL_i/H_d)$ & --2 &--2 & $-16$ \\
\hline
 \bf      1    &   ${\bf 1}~ (XN_i/S)$                  &  0 & 4 & $-2$ \\
\hline
\end{tabular}
\end{center}
\caption{Decomposition of the $E_6$ fundamental  ${\bf 27}$
representation under $SO(10)$, $SU(5)$, and the $U(1)_{\chi}$,
$U(1)_{\psi}$ and $U(1)'$ charges  of multiplets.
The SM quark doublets, right-handed  up-type quarks, right-handed down-type quarks, lepton doublets, right-handed charged leptons,
and right-handed neutrinos are labeled as $Q_i$, $U_i^c$, $D_i^c$, $L_i$, $E_i^c$, and $N_i^c$, respectively.}
\label{E6charge}
\end{table}

In our model we introduce three fermionic ${\bf 27}$s, two pairs of vector-like fermions,
one scalar Higgs doublet field $H_u$ from the doublet of ${\bf {5}}$ of $SU(5)$,
one scalar Higgs doublet field $H_d$ from the doublet of ${\bf {\bar 5}}$ of $SU(5)$, 
one scalar SM singlet Higgs field $T$ from the singlet of ${\bf 16}$ of $SO(10)$,
and one scalar SM singlet Higgs field $S$ from the singlet of ${\bf 27}$ of $E_6$. 
 In particular, to realize the isospin violation, we assume that 
the first two generations of the left-handed quark doublets $Q_k$  have $U(1)'$ charge $9$,
while $Q_3$ and $XQ_k$ have $U(1)'$ charge $-9$, where $k=1,~2$. To cancel the gauge anomaies, 
we introduce $\overline{XQ}_k$. 
Note that the additional fermions from the ${\bf 27}$ with masses at the TeV 
scale are
%In addition, the fermions in the ${\bf 27}$ representations are
 $N_i^c$, $XD_i$, $XL_i^c$, $XD_i^c$, $XQ_k$, $\overline{XQ}_k$, $XL_i$, and $XN_i$. 
 For details, please see 
 Table \ref{Particle-Spectrum}.

By choosing 
\begin{eqnarray}
\tan\theta = -{1\over {17}}\sqrt{3/5}~,~\,
\end{eqnarray}
it is natural to realize IVDM with $f_n/f_p=-0.7$.

\begin{table}[h]
\begin{tabular}{|c|c|c|c|c|c|}
\hline
$Q_k$ &~$(\mathbf{3}, \mathbf{2}, \mathbf{1/6}, \mathbf{9})$~ &
~$Q_3/XQ_k$~ & ~$(\mathbf{3}, \mathbf{2}, \mathbf{1/6}, \mathbf{-9})$~ &
~$\overline{XQ}_k$~ & ~$(\mathbf{\overline{3}}, \mathbf{2}, \mathbf{-1/6}, \mathbf{-9})$ ~\\
\hline
$U_i^c$ &  ~$(\mathbf{\overline{3}}, \mathbf{1}, \mathbf{-2/3}, \mathbf{-9})$~ &
~$D_i^c$~ & ~$(\mathbf{\overline{3}}, \mathbf{1}, \mathbf{1/3}, \mathbf{25})$ ~&
~$L_i$~ & ~$(\mathbf{1}, \mathbf{2},  \mathbf{-1/2}, \mathbf{25})$~\\
\hline
$E_i^c$ &  $(\mathbf{1}, \mathbf{1},  \mathbf{1}, \mathbf{-9})$ &
~$N_i^c/T$~ &  $(\mathbf{1}, \mathbf{1},  \mathbf{0}, \mathbf{-43})$ &
~$XD_i$~ & ~$(\mathbf{3}, \mathbf{1}, \mathbf{-1/3}, \mathbf{18})$~  \\
\hline
~$XL^c_i,~H_u$~ & ~$(\mathbf{1}, \mathbf{2},  \mathbf{1/2}, \mathbf{18})$~ &
~$XD_i^c$~ & ~$(\mathbf{\overline{3}}, \mathbf{1}, \mathbf{1/3}, \mathbf{-16})$ &
~$XL_i,~H_d$~ & ~$(\mathbf{1}, \mathbf{2},  \mathbf{-1/2}, \mathbf{-16})$~ \\
\hline
~$XN_i,~S$~ &  ~$(\mathbf{1}, \mathbf{1},  \mathbf{0}, \mathbf{-2})$~ & 
~$\chi$~ &  ~$(\mathbf{1}, \mathbf{1},  \mathbf{0}, \mathbf{-27/2})$~ &
~$\Phi$~&  ~$(\mathbf{1}, \mathbf{1},  \mathbf{0}, \mathbf{86})$~   \\
\hline
~$S'$~ & ~$(\mathbf{1}, \mathbf{1},  \mathbf{0}, \mathbf{18})$~ &
~$\phi$ & ~$(\mathbf{1}, \mathbf{1},  \mathbf{0}, \mathbf{4})$~ &
$\Phi'$ & ~$(\mathbf{1}, \mathbf{1},  \mathbf{0}, \mathbf{1})$~  \\
\hline
~$\phi'$~&  ~$(\mathbf{1}, \mathbf{1},  \mathbf{0}, \mathbf{7})$~   
 & ~$\phi''$ & ~$(\mathbf{1}, \mathbf{1},  \mathbf{0}, \mathbf{5})$~     & &\\
\hline
\end{tabular}
\caption{The quantum number assignment for particles under
  $SU(3)_C \times SU(2)_L \times U(1)_Y \times U(1)'$ gauge symmetry, $i=1,~2,~3$, and $k=1,~2$. Here,
  the correct $U(1)'$ charges are the $U(1)'$ charges in the Table divided
  by $4{\sqrt{181}}$.}
\label{Particle-Spectrum}
\label{tab:charge}
\end{table}

Three SM singlet Higgs fields $\Phi$, $\phi$, and $S'$ with $U(1)'$ charges ${\bf 86}$, ${\bf 4}$, and ${\bf 18}$ are introduced to 
generate the masses for $N_i^c$, $XN_i$, and vector-like fermions ($XQ_k$, $\overline{XQ}_k$),
respectively.
In order to break the global symmetries in the Higgs potential and avoid the massless Nambu-Glodstone bosons, 
we introduce two SM singlet Higgs fields $\Phi'$, $\phi'$, 
and $\phi''$ with $U(1)'$ charges ${\bf 1}$, ${\bf 7}$, and ${\bf 5}$, 
respectively. Moreover, to introduce a dark matter candidate, we introduce 
a SM singlet Dirac fermion $\chi$ with $U(1)'$ charge ${\bf -27/2}$, 
which will not affect the gauge anomaly cancellations.
In particular, only the $U(1)'$ charge of $\chi$ is a half integer while the $U(1)'$ charges of 
 all the other particles are integers.
And then after $U(1)'$ gauge symmetry breaking, there exists a residual discrete $Z_2$ gauge symmetry
under which $\chi$ is odd while all the other particles are even.
Thus, $\chi'$ cannot decay and can be a dark matter candidate.
For details, please see Table \ref{Particle-Spectrum} as well.

The interesting question is whether $\chi$, ($Q_K$, $\overline{XQ}_k$), $\Phi$, $S'$, $\phi$,  $\Phi'$, 
$\phi'$, and $\phi''$ can arise from the higher representations of $E_6$ since they do not belong
to the fundamental ${\bf 27}$ representation of $E_6$. And let us discuss it one by one.

First, $\chi$ cannot arise from the $E_6$ representations since it is stable and cannot decay due to 
the discrete $Z_2$ gauge symmetry from $U(1)'$ gauge symmetry breaking. Also, $\chi$ is a Dirac fermion
and then might  play a role of asymmetric dark matter which can affect the calculation of 
the dark matter relic density. This is very interesting, but for simplicity, we shall not consider
it here.

Second, to address whether ($Q_K$, $\overline{XQ}_k$) can arise from the higher representations of $E_6$,
we only need to consider $Q_k$ for simplicity. We can list all the particles from the higher representations of $E_6$
whose quantum numbers under $SU(3)_C\times SU(2)_L\times U(1)_Y \times U(1)'$ 
are $(\mathbf{3}, \mathbf{2}, \mathbf{1/6}, \mathbf{Q'})$ as follows
\begin{eqnarray}
&& (\mathbf{3}, \mathbf{2}, \mathbf{1/6}, \mathbf{-54}) \subset \mathbf{2925} ~{\rm of}~E_6~,~ \nonumber \\
&& (\mathbf{3}, \mathbf{2}, \mathbf{1/6}, \mathbf{-52}) \subset \overline{\mathbf{351'}} ~{\rm of}~E_6~,~ \nonumber \\
&&  (\mathbf{3}, \mathbf{2}, \mathbf{1/6}, \mathbf{-50}) \subset \mathbf{351},~{\rm and}~ \mathbf{1728} ~{\rm of}~E_6~,~ \nonumber \\
&& (\mathbf{3}, \mathbf{2}, \mathbf{1/6}, \mathbf{-48}) \subset \mathbf{2430},~{\rm and}~ \mathbf{2925} ~{\rm of}~E_6~,~ \nonumber \\
&& (\mathbf{3}, \mathbf{2}, \mathbf{1/6}, \mathbf{-11}) \subset \overline{\mathbf{351'}},
~{\rm and}~ \overline{\mathbf{1728}} ~{\rm of}~E_6~,~ \nonumber \\
&& (\mathbf{3}, \mathbf{2}, \mathbf{1/6}, \mathbf{-10}) \subset \mathbf{650} ~{\rm of}~E_6~,~ \nonumber \\
&& (\mathbf{3}, \mathbf{2}, \mathbf{1/6}, \mathbf{-9}) \subset \mathbf{27},~\mathbf{351},~\mathbf{351'},
~{\rm and}~\mathbf{1728}~{\rm of}~E_6~,~ \nonumber \\
&& (\mathbf{3}, \mathbf{2}, \mathbf{1/6}, \mathbf{-7}) \subset \mathbf{78},~\mathbf{650},~\mathbf{2430},
~{\rm and}~\mathbf{2925}~{\rm of}~E_6~,~ \nonumber \\
&&  (\mathbf{3}, \mathbf{2}, \mathbf{1/6}, \mathbf{32}) \subset \mathbf{351},~{\rm and}~ \mathbf{1728} ~{\rm of}~E_6~,~ \nonumber \\
&& (\mathbf{3}, \mathbf{2}, \mathbf{1/6}, \mathbf{34}) \subset \mathbf{78},~\mathbf{650},~\mathbf{2430},
~{\rm and}~\mathbf{2925}~{\rm of}~E_6~,~ \nonumber \\
&& (\mathbf{3}, \mathbf{2}, \mathbf{1/6}, \mathbf{75}) \subset \mathbf{2430},
~{\rm and}~\mathbf{2925}~{\rm of}~E_6~,~ \nonumber \\
&& (\mathbf{3}, \mathbf{2}, \mathbf{1/6}, \mathbf{77}) \subset \overline{\mathbf{1728}} ~{\rm of}~E_6~.~\,
\end{eqnarray}
Therefore,  ($Q_K$, $\overline{XQ}_k$) cannot arise from the higher representations of $E_6$.

Third, to address whether the scalars $\Phi$, $S'$, $\phi$, $\Phi'$, $\phi'$, and $\phi''$ can arise from the higher representations of $E_6$,
we present all the particles from the higher representations of $E_6$
whose quantum numbers under $SU(3)_C\times SU(2)_L\times U(1)_Y \times U(1)'$ 
are $(\mathbf{1}, \mathbf{1}, \mathbf{0}, \mathbf{Q'})$ as follows
\begin{eqnarray}
&& (\mathbf{1}, \mathbf{1}, \mathbf{0}, \mathbf{-86}) \subset \overline{\mathbf{351'}} ~{\rm of}~E_6~,~ \nonumber \\
&& (\mathbf{1}, \mathbf{1}, \mathbf{0}, \mathbf{-84}) \subset  \mathbf{1728} ~{\rm of}~E_6~,~ \nonumber \\
&& (\mathbf{1}, \mathbf{1}, \mathbf{0}, \mathbf{-82}) \subset  \mathbf{2430} ~{\rm of}~E_6~,~ \nonumber \\
&& (\mathbf{1}, \mathbf{1}, \mathbf{0}, \mathbf{-45}) \subset \overline{\mathbf{351'}} ~{\rm of}~E_6~,~ \nonumber \\
&& (\mathbf{1}, \mathbf{1}, \mathbf{0}, \mathbf{-44}) \subset \mathbf{650} ~{\rm of}~E_6~,~ \nonumber \\
&& (\mathbf{1}, \mathbf{1}, \mathbf{0}, \mathbf{-43}) \subset \mathbf{27},~\mathbf{351},
~{\rm and}~\mathbf{1728}~{\rm of}~E_6~,~ \nonumber \\
&& (\mathbf{1}, \mathbf{1}, \mathbf{0}, \mathbf{-41}) \subset \mathbf{78},~\mathbf{2430},
~{\rm and}~\mathbf{2925}~{\rm of}~E_6~,~ \nonumber \\
&& (\mathbf{1}, \mathbf{1}, \mathbf{0}, \mathbf{-39}) \subset \overline{\mathbf{1728}} ~{\rm of}~E_6~,~ \nonumber \\
&& (\mathbf{1}, \mathbf{1}, \mathbf{0}, \mathbf{-4}) \subset \overline{\mathbf{351'}} ~{\rm of}~E_6~,~ \nonumber \\
&& (\mathbf{1}, \mathbf{1}, \mathbf{0}, \mathbf{-2}) \subset \mathbf{27}, ~\mathbf{351},~{\rm and}~\mathbf{1728} ~{\rm of}~E_6~.~\,
\end{eqnarray}
The singlet scalar particles with positive charges can be obtained from above particles via Hermitian conjugage.
Therefore, $\Phi$ and $\phi$ can arise from the $\mathbf{351'}$ representation of $E_6$,
while  $S'$, $\Phi'$, $\phi'$, and $\phi''$ cannot arise from the higher representations of $E_6$.

Interestingly, ($Q_K$, $\overline{XQ}_k$),  $S'$, $\Phi'$, $\phi'$, and $\phi''$ might emerge
as the composite states. For example,  $Q_K$ can arise from 
$(\mathbf{78/650/2430/2925}\times \mathbf{351'}) \times \mathbf{351'} \times \mathbf{351'} \times \mathbf{351'} $,  
$\overline{XQ}_k$ can arise from the Hermitian conjugates of the above terms,
$S'$ can arise from 
$(\mathbf{\overline{27}/\overline{351}/\overline{1728}})\times \mathbf{351'} 
\times \mathbf{351'} \times \mathbf{351'} \times \mathbf{351'}$,
$\Phi'$ can arise from 
$\mathbf{351'} \times \mathbf{650}$ or $\mathbf{\overline{650}} \times (\mathbf{27/351/1728})$,
$\phi'$ can arise from 
$\mathbf{351'} \times \mathbf{\overline{650}} \times  (\mathbf{78/2430/2925})$,
and $\phi''$ can arise from 
$\mathbf{\overline{650}}  \times \mathbf{\overline{1728}}$.

The Higgs potential for the $U(1)'$ gauge symmetry breaking is 
\begin{eqnarray}
  V &=& -m_S^2 |S|^2 -m_T^2 |T|^2 -m_{\Phi}^2 |\Phi|^2 -m_{S'}^2 |S'|^2 -m_{\phi}^2 |\phi|^2  -m_{\phi'}^2 |\phi'|^2 -m_{\Phi'}^2 |\Phi'|^2 - -m_{\phi''}^2 |\phi''|^2
\nonumber\\&&
+ \lambda_S |S|^4 
+ \lambda_T |T|^4
+ \lambda_\Phi |\Phi|^4 +  \lambda_{S'} |S'|^4 + \lambda_\phi |\phi|^4 + \lambda_\phi' |\phi'|^4 
+ \lambda_\Phi' |\Phi'|^4 
+ \lambda_{\phi''} |\phi''|^4 
\nonumber\\&&
+ \lambda_{ST} |S|^2|T|^2 
+ \lambda_{S\Phi} |S|^2|\Phi|^2 
+ \lambda_{SS'} |S|^2|S'|^2 +\lambda_{S\phi} |S|^2|\phi|^2 +\lambda_{S\phi'} |S|^2|\phi'|^2 +\lambda_{S\Phi'} |S|^2|\Phi'|^2
\nonumber\\&&
+\lambda_{S\phi''} |S|^2|\phi''|^2+
+ \lambda_{T\Phi} |T|^2|\Phi|^2 + \lambda_{TS'} |T|^2|S'|^2 +\lambda_{T\phi} |T|^2|\phi|^2 
+\lambda_{T\phi'} |T|^2|\phi'|^2 +\lambda_{T\Phi'} |T|^2|\Phi'|^2
\nonumber\\&&
+\lambda_{T\phi''} |T|^2|\phi''|^2
+\lambda_{\Phi S'} |\Phi|^2|S'|^2 +
\lambda_{\Phi\phi} |\Phi|^2|\phi|^2 +\lambda_{\Phi\phi'} |\Phi|^2|\phi'|^2 +\lambda_{\Phi\Phi'} |\Phi|^2|\Phi'|^2 +
\lambda_{\Phi\phi''} |\Phi|^2|\phi''|^2 
\nonumber\\&&
+\lambda_{S'\phi} |S'|^2|\phi|^2
+\lambda_{S'\phi'} |S'|^2|\phi'|^2
+\lambda_{S'\Phi} |S'|^2|\Phi|^2
+\lambda_{S'\phi''} |S'|^2|\phi''|^2
+\lambda_{\phi\phi'} |\phi|^2|\phi'|^2 +\lambda_{\phi\Phi'} |\phi|^2|\Phi'|^2 
\nonumber\\&&
+\lambda_{\phi\phi''} |\phi|^2|\phi''|^2 
+\lambda_{\phi'\Phi'} |\phi'|^2|\Phi'|^2 
+\lambda_{\phi'\phi''} |\phi'|^2|\phi''|^2 
+\lambda_{\Phi'\phi''} |\Phi'|^2|\phi''|^2 
\nonumber\\&&
+\left(A_1 S H_d H_u+ A_2 T^2 \Phi
+A_3 S (\Phi')^2  +A_4 S^2\phi +A_5 S \phi' (\phi'')^{\dagger}+ A_6 \phi \Phi' (\phi'')^{\dagger}
\right. \nonumber\\ && \left. 
+  \lambda_1 (\phi^2)^{\dagger} \phi' \Phi' 
+ \lambda_2 (S')^2 \phi' T + \lambda_3 \Phi'^2 (H_dH_u)^{ \dagger}
+\lambda_4 \phi'' S^2 (\Phi')^{\dagger}+ \lambda_5 (\phi')^{\dagger}  \phi'' (\Phi')^2
\right. \nonumber\\ && \left. 
+ \lambda_6 \phi' (\phi'')^{\dagger} (H_dH_u)^{ \dagger} 
+ \lambda_7 S^{\prime \dagger}(\phi')^2 \phi+{\rm H. c.}~\right) ~.~\,
\end{eqnarray}
Note that without the $A_i$ and $\lambda_i$ terms in the bracket, there are 
ten global $U(1)$ symmetries for the field $S$, $T$, $\Phi$, $S'$, $\phi$, $\Phi'$,
$\phi'$, $\phi''$, $H_u$, and $H_d$. After they obtain the Vacuum Expectation Values (VEVs), 
we have ten Goldstone bosons, where two of them are eaten by $U(1)_Y$ and the extra $U(1)'$ gauge boson. 
Thus, to avoid the extra Goldstone bosons, we need the $A_i$ and $\lambda_i$ terms to break eight 
global $U(1)$ symmetries.
 By the way,  $\lambda_1$ term can be regarded 
as the combination of $A_3$, $A_4$, $A_5$, and $A_6$ terms,
 $\lambda_3$ term can be regarded as the combination of $A_1$ and $A_3$ terms,
$\lambda_4$ term can be regarded as the combination of $A_4$ and $A_6$ terms,
$\lambda_5$ term can be regarded as the combination of $A_3$ and $A_5$ terms,
and $\lambda_6$ term can be regarded as the combination of $A_1$ and $A_5$ terms. 
Thus, we are left with only two global symmetries in the above potential, 
which are $U(1)_Y$ and the extra $U(1)'$ gauge symmetry. 
Also, $S$, $T$, $\Phi$, $S'$, $\phi$, $\Phi'$, $\phi'$, and $\phi''$ will mix with each other via the quartic  
and trilinear terms.
In addition, the $U(1)'$ symmetry breaking Higgs fields $S$, $T$, $\Phi$, $S'$, $\phi$, $\Phi'$, 
$\phi'$, $\phi''$ and the
electroweak symmetry breaking Higgs fields $H_u$ and $H_d$ can be mixed via the quartic terms as well,
for example, $|S|^2 |H_u|^2$, etc, which can be written down easily. For simplicity,
we will not study it here, and neglect the mass terms and quartic terms for $H_u$ and $H_d$ as well.

The Yukawa couplings in our models are 
\begin{eqnarray}
  -{\cal L} &=& y_{3i}^U Q_3 U_i^c H_u + y_{ki}^U XQ_k U_i^c H_u 
+  y_{3i}^D Q_3 D_i^c H_d + y_{ki}^D XQ_k D_i^c H_d 
 + y^{XQ}_{k3} S' \overline{XQ}_k Q_3
\nonumber\\&&
+ y^{XQ}_{kl} S' \overline{XQ}_k XQ_l
+ y_{ij}^E L_i E_j^c H_d
  + y_{ij}^N L_i N_j^c H_u + y_{ij}^{XNd} XL_i^c XN_j H_d \nonumber\\&&
  + y_{ij}^{XNu} XL_i XN_j H_u 
  + y_{ij}^{TD} D_i^c XD_j {T} + y_{ij}^{TL} XL_i^c L_j {T}
+ y_{ij}^{SD} XD_i^c XD_j S
\nonumber\\&&
  + y_{ij}^{SL} XL_i^c XL_j S + y_{ij}^{N^c} N_i^c N^c_j \Phi + y_{ij}^{XN} XN_i XN_j \phi
  +M^D_{\chi}{\bar \chi} \chi + M_{kl} \overline{XQ}_k Q_l + {\rm H. c.}~,~\,    
\end{eqnarray}
where $i,~j=1,~2,~3$, and $k,~l=1,~2$.
Thus,  after $S$, $T$, and $S'$ obtain VEVs or after $U(1)'$ gauge symmetry breaking,
$(XD_i^c,~XD_i)$ and $(XL_i^c,~XL_i)$ will become vector-like particles
from the $y_{ij}^{SD} XD_i^c XD_j S$ and $y_{ij}^{SL} XL_i^c XL_j S$ terms,
 $(D_i^c,~XD_i)$ and $(XL_i^c,~L_i)$ will obtain vector-like masses
from the $ y_{ij}^{TD} D_i^c XD_j {T}$ and $y_{ij}^{TL} XL_i^c L_j {T}$ terms,
and $(XQ_k,~\overline{XQ}_k)$ will obtain vector-like masses from the
$y^{XQ}_{kl} S' \overline{XQ}_k XQ_l$ terms.
After diagonalizing their mass matrices, we obtain the mixings
between $XD_i^c$ and $D_i^c$, and the mixings between $XL_i$ and $L_i$.
The discussion of the Higgs potential for electroweak symmetry breaking is similar to
the Type II two Higgs doublet model, so we will not repeat it here.
In addition, the third-generation quark masses are obtained directly from 
$y_{33}^U Q_3 U_3^c H_u $ and $y_{33}^D Q_3 D_3^c H_d$ terms,
while the first two-generation quark masses are obtained by integrating out the
vector-like particles $(XQ_k,~\overline{XQ}_k)$ after $U(1)'$ gauge symmetry breaking.
Thus, we can explain why the first two-generation quarks are lighter than the third-generation in our model.

At low energy, the relevant  degrees of freedom are SM particles, $Z'$, and DM $\chi$.
The interactions can be expressed as
\begin{eqnarray}
-{\cal L} = \sum_q g_{u} \bar{u} \gamma^{\mu}uZ_{\mu}^\prime + g_{uA} \bar{u} \gamma^{\mu}\gamma^{5}uZ_{\mu}^\prime + g_{d} \bar{d} \gamma^{\mu}dZ_{\mu}^\prime + g_{dA} \bar{d} \gamma^{\mu}\gamma^{5}dZ_{\mu}^\prime  +g_{\chi} \bar{\chi} \gamma^{\mu}\chi Z_{\mu}^\prime.
\end{eqnarray}
The ratio of different $U(1)'$ couplings are determined by its $U(1)'$ charge tabulated in Table~\ref{tab:charge}.
Since our model is isospin-violated,  $u$ and $d$ quarks couple different with $Z'$. 
After a brief combination, we  get
\begin{equation}
g_u:g_d:g_\chi :g_{dA}=18:-16:-27:-34,~g_{uA}=0.
\end{equation}
For example, if we set $g_{u} = 0.1$ , then $g_{d} = -0.0889$ , $g_{\chi} = -0.15$ , $g_{uA} = 0$ , $g_{dA} = -0.1889$. Particularly, we do not have axial vector terms for $u$-quark in our model.

\section{Constraints from dark matter experiments}
\label{sec:DD}
%By definition, isospin-violating dark matter is a generic framework that includes all dark matter candidates that interact differently with protons and neutrons.
% The first example of such a dark matter particle to have been studied was likely the heavy Dirac neutrino, which couples more strongly to the neutrons than to the protons. 

Generally, DM direct detection experiments assume DM couples the same to proton and neutron, and then report their limits for cross sections per nucleon.
In the more general framework of IVDM, the cross sections per nucleon $ \sigma_N^Z$ is defined as
\begin{equation}
\sigma_N^Z = \sigma_p \frac{\Sigma_i \eta_i \mu_{A_i}^2 [Z+(A_i-Z)f_n/f_p]^2}{\Sigma_i \eta_i \mu^2_{A_i}A_i^2} \equiv \frac{ \sigma_p}{F_Z},
\end{equation}
where $A_i$ refers to different isotopes and $\eta_i$ is corresponding fractional number abundance.
If $\tilde{\sigma}$ is the limit reported  by an experiment, then $F_Z \tilde{\sigma}$ is the limit for IVDM.
It is obvious that the DM elastic scattering off nucleus will have coherent effect between $\sigma_p$ and $\sigma_n$, which can leads to a strongly destructive effect with particular $f_n/f_p$.

\begin{figure}[ht]
	\centering
	\begin{subfigure}{0.45\textwidth} % width of left subfigure
		\includegraphics[width=\textwidth]{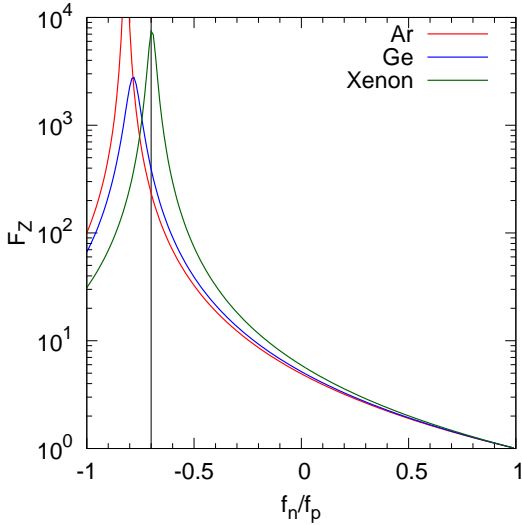}
		\caption{$F_Z$ for Ar/Ge/Xenon} % subcaption
	\end{subfigure}
	\hspace{3em} % here you can insert horizontal or vertical space
	\begin{subfigure}{0.45\textwidth} % width of right subfigure
		\includegraphics[width=\textwidth]{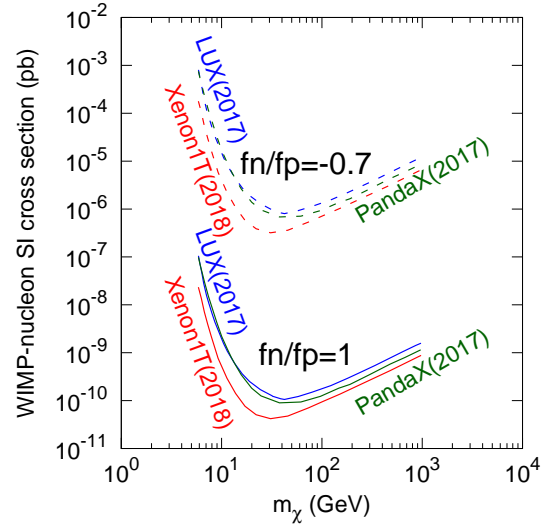}
		\caption{The rescale of the IVDM limit.} % subcaption
	\end{subfigure}
	\caption{The scaling factor $F_Z$ for three different materials (left) and the rescaled limits of three Xenon  based experiment (right), e.g., PandaX-II (2017)~\cite{Cui:2017nnn}, LUX (2017)~\cite{Akerib:2016vxi},  and Xenon1T (2018)~\cite{Aprile:2018dbl}.} % caption for whole figure
	\label{FZ}
\end{figure}

Previously, the IVDM with new experimental data has been studied in Ref.~\cite{Yaguna:2016bga}. 
Here we update some experiment results and apply this bounds to our model. 
Shown in the left panel of Fig.~\ref{FZ} are $F_Z$ for three kinds of materials with isotopy effects taken into account. 
For the case of Xenon, $F_Z$ get its maximum at $f_n/f_p = -0.7$. 
In the right panel of Fig.~\ref{FZ} we present the rescaled limits for three kinds of direct detection experiments.
It is obvious that constraints of these Xenon based experiments could be relaxed by a factor of about $10^{-4}$. 
 
\begin{figure}[htb!]
	\centering
\includegraphics[scale=0.9]{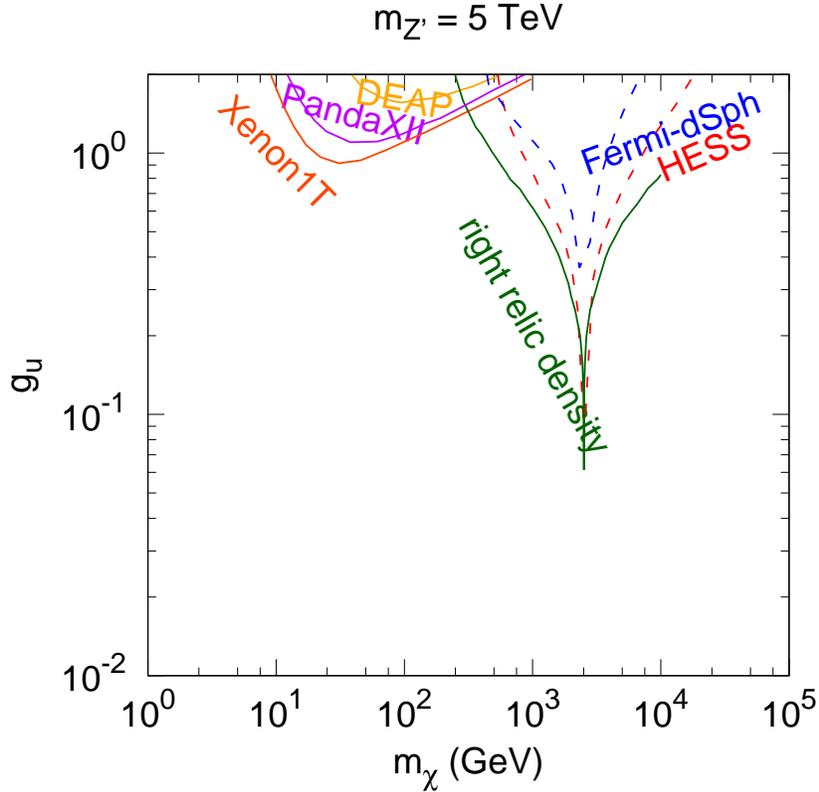}
%	\begin{subfigure}{0.45\textwidth} % width of left subfigure
%		\includegraphics[width=\textwidth]{1}
%		\caption{$m_{Z'} = 1$ TeV}  % subcaption
%	\end{subfigure}
%	\hspace{3em} % here you can insert horizontal or vertical space
%	\begin{subfigure}{0.45\textwidth} % width of right subfigure
%		\includegraphics[width=\textwidth]{5}
%		\caption{$m_{Z'} = 5$ TeV} % subcaption
%	\end{subfigure}
	\caption{Estimated $90\%$ C.L. limits in $m_{\chi}-g_u$ plane for  direct detection experiments and indirect detection experiments. The solid orange, purple, and yellow lines  correspond to PandaX-II (2017)~\cite{Cui:2017nnn}, Xenon1T (2018)~\cite{Aprile:2018dbl}, DEAP3600 (2017)~\cite{Amaudruz:2017ekt} experiments, respectively. The dashed blue and red lines correspond to Fermi-dSph (6-year)~\cite{TheFermi-LAT:2017vmf} and HESS (254h)~\cite{Abdallah:2016ygi}, respectively. 
The dark-green line indicates the parameter space with the observed dark matter relic density.}. % caption for whole figure
\label{mxgu}
\end{figure}

\begin{figure}[htb!]
	\centering
	\begin{subfigure}{0.45\textwidth} % width of left subfigure
		\includegraphics[width=\textwidth]{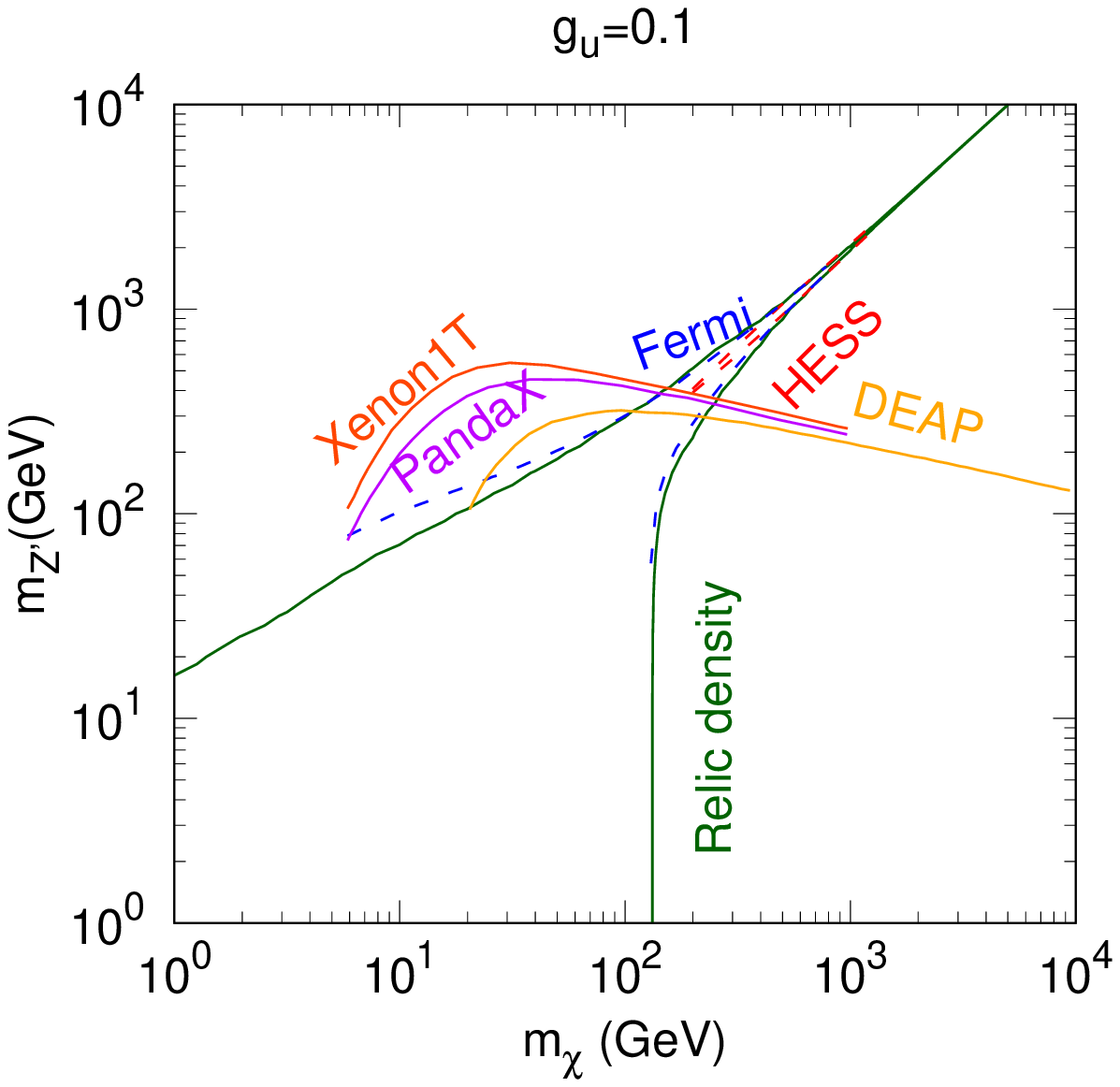}
		\caption{$g_u$ = 0.1} % subcaption
	\end{subfigure}
	\hspace{3em} % here you can insert horizontal or vertical space
	\begin{subfigure}{0.45\textwidth} % width of right subfigure
		\includegraphics[width=\textwidth]{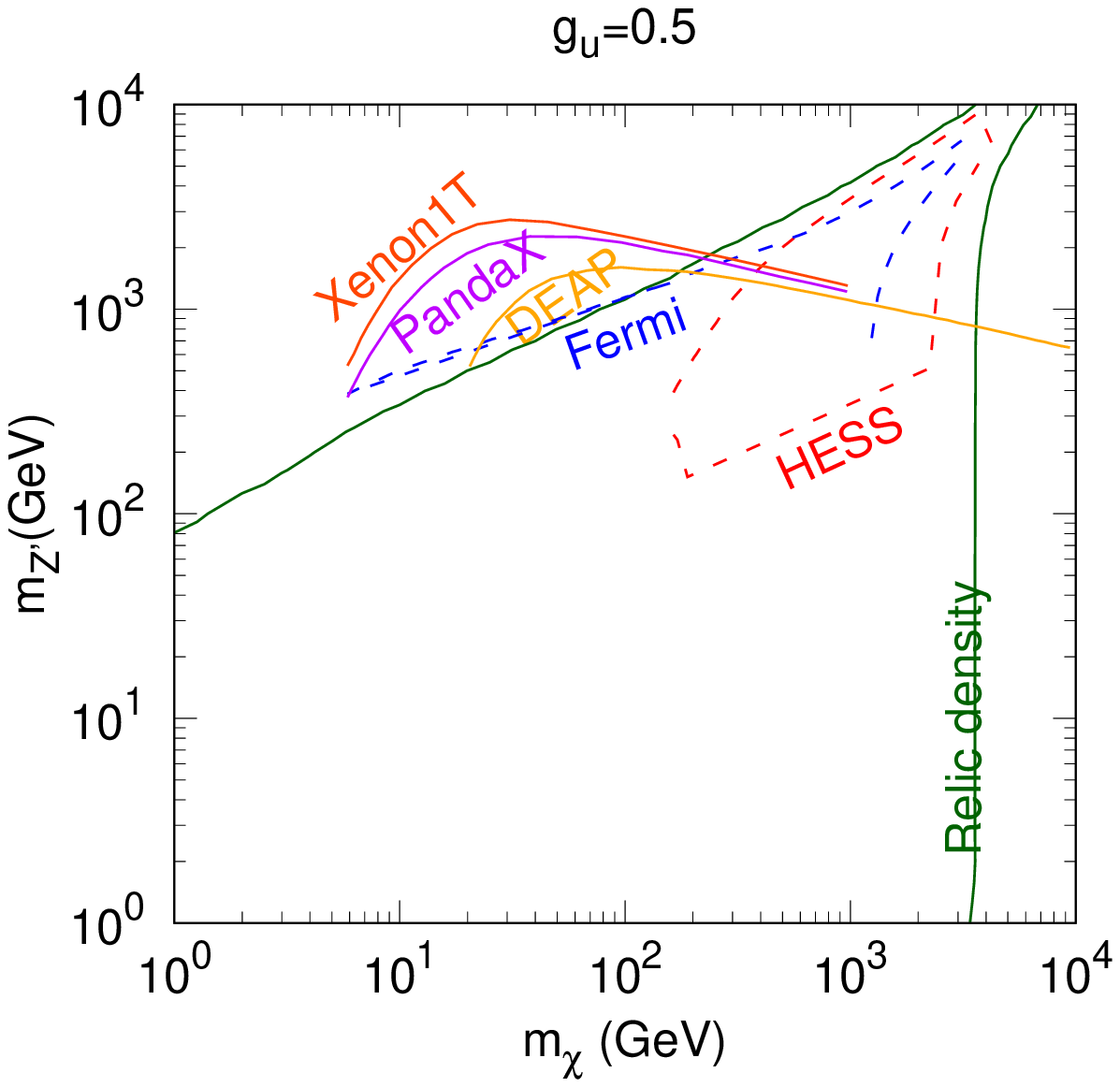}
		\caption{$g_u$ = 0.5} % subcaption
	\end{subfigure}
	
	\caption{The estimated $90\%$ C.L. limits in $m_{\chi}-m_{Z'}$ plane with the meaning of lines  are the same in Fig.~\ref{mxgu}
.} % caption for whole figure
	\label{mxmz}
\end{figure}

It is well-known that for scalar and vector interaction, direct detection experiments have stronger capability to detect heavy DM with masses larger than~10 GeV, while collider searches have better sensitivity for small DM~\cite{Alves:2013tqa,Buchmueller:2014yoa,Abdallah:2014hon,Xiang:2015lfa}.
This conclusion would change dramatically once the isospin-violating effects are taken into account.
In Figs.~\ref{mxgu} and \ref{mxmz} we show the limits from direct detection experiment and indirect detection experiments. 
It is obvious that near the region of $m_\chi \sim \frac{1}{2}m_{Z'} $, the line of 
the correct DM relic density varies sharply due to resonant enhancement.
Aside from the resonance region, DM direct detection experiments have better sensitivities than DM indirect detection experiments; while the latter give the best sensitivity around $m_\chi \sim \frac{1}{2}m_{Z'}$.
In Figs.~\ref{mxgu} and \ref{mxmz}, we also demonstrate the region satisfying the observed relic density, which roughly trace the sensitivities of indirect detection experiment due to the $s$-wave annihilation nature of DM, as shown in the Appendix~\ref{app:anni}.

\section{Constraints from Future Collider}
\label{sec:collider}

Another powerful methods to explore the nature of DM is collider search.
In our model DM interacts directly with quarks, and can be copiously produced at hadron colliders such as the LHC  and proposed LHC-hh~\cite{Benedikt:2018csr} and SppC~\cite{CEPC-SPPCStudyGroup:2015csa}.
Once DM are produced, they will escape the detectors undetected, so another additional radiation is
 needed to trace these events. 
 In this section we study the  sensitivities of future colliders for this model, and compare them with 
those obtained from DM direct and indirect experiments.
The techniques of collider research closely follow Ref.~\cite{Xiang:2015lfa}.

\begin{figure}[ht]
	\centering
	\begin{subfigure}{0.45\textwidth} % width of left subfigure
		\includegraphics[width=1.0\textwidth]{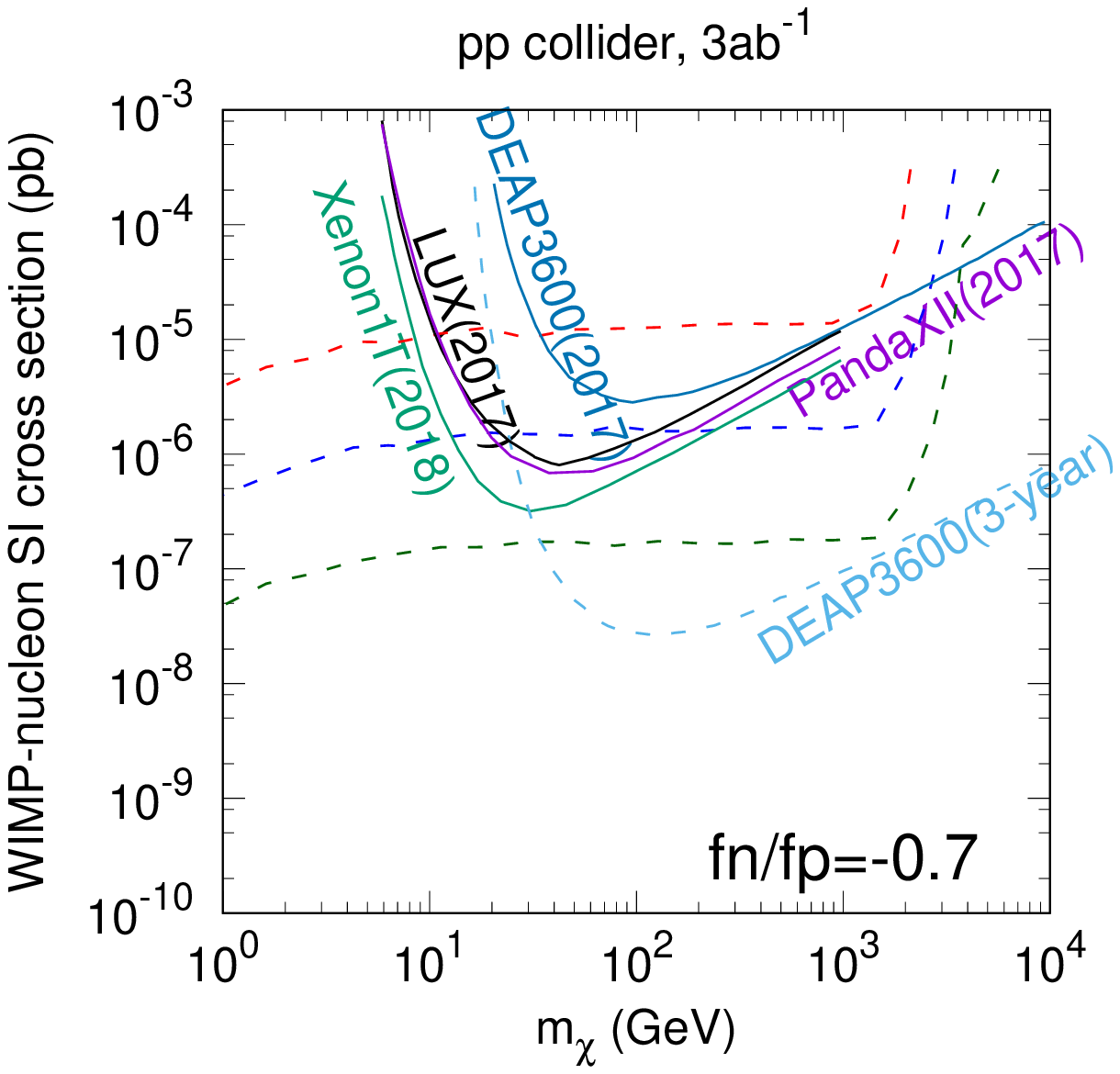}
		\caption{Direct detection vs future collider } % subcaption
	\end{subfigure}
	\hspace{3em} % here you can insert horizontal or vertical space
	\begin{subfigure}{0.45\textwidth} % width of right subfigure
		\includegraphics[width=1.0\textwidth]{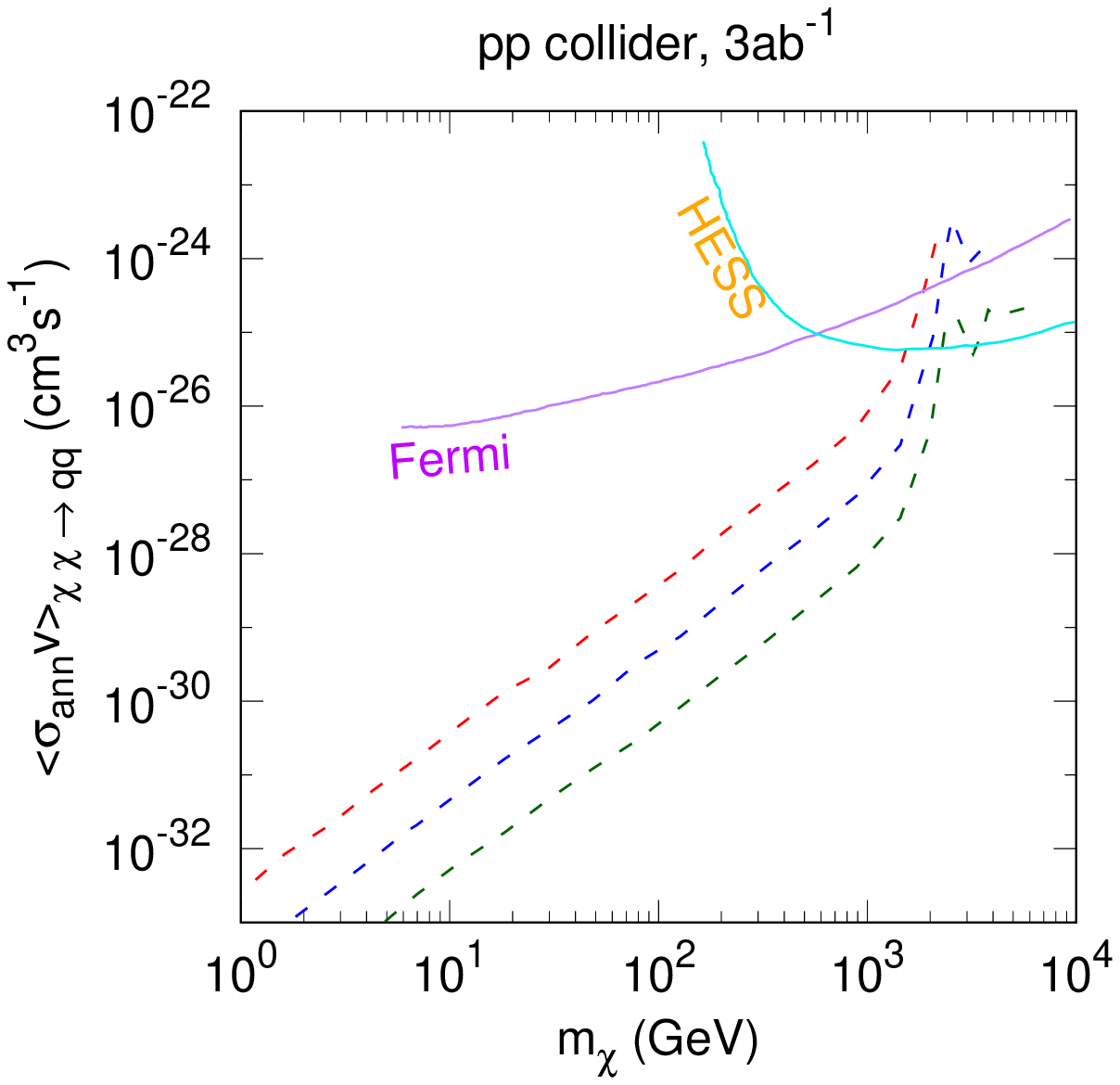}
		\caption{Indirect detection vs future collider} % subcaption
	\end{subfigure}
	
	\caption{The estimated  limits  for  different detection methods. The red, blue, and dark-green lines correspond
 to future colliders with energy at $\sqrt{s} = $33, 50, and 100~TeV, respectively.  
The  dashed lines correspond to the benchmark choice with $m_{Z^\prime}$ equals to 5 TeV.} 
	\label{futurecollider}
\end{figure}

In this  study, we focus on the monojet signal process $pp \rightarrow Z^{\prime(*)} \rightarrow \chi \bar{\chi} +$ jets.
The main backgrounds are $Z(\rightarrow \bar{\nu}\nu) + $jets and $W(\rightarrow l \nu) +$ jets.
Background and signal events at the parton level are generated with \texttt{MadGraph 5}~\cite{Alwall:2014hca} and then we use \texttt{PYTHIA~8}~\cite{Sjostrand:2007gs} to do parton shower and hadronization.
MLM matching scheme are chose to avoid events double counting from matrix calculation and parton shower.
We adopt \texttt{Delphes~3}~\cite{deFavereau:2013fsa} to perform  fast detector simulation.
Jets are reconstructed with anti-$K_T$ algorithm with a distance parameter $R=0.4$.
 The future colliders would be constructed with higher resolution, so the results here are conservative 
and expected to be improved.

%\begin{table}[ht]
%	\centering
%	\caption{Cut for matching}
%	\begin{tabular}{c c c c}
%		\hline
%		\hline
%		$\sqrt{s}$  & xqcut  & ptj1min   & Cut on ptj1 ($\slashed{E}_T, P_T)$ \\
%		\hline
%		33 TeV & 30   GeV & 800 GeV  & $1600 $ GeV  \\
%		\hline
%		50 TeV & 50 GeV & 1000 GeV & 1800 GeV   \\
%		\hline
%		100 TeV & 70 GeV & 1600 GeV & 2600 GeV \\
%		\hline
%		\hline
%	\end{tabular}
%	\label{matching}
%\end{table}

To improve the statistical significance, several cuts are implemented on both signal  and background events.
There must be at least two energetic jet in the final states. 
The leading jet $j_1$ is required to have $|\eta(j_1)| < 2.4$ and $p_T(j_1) >$  1.6/1.8/2.6 TeV for $\sqrt{s}=$ 33/50/100 TeV. 
Events with more than two jets with $p_T>$ 100 GeV and $|\eta|<$ 4 are rejected. The DM production process may involve more than one jet from initial state radiation. In order to keep more signal events, a second jet($j_2$) is allowed if it satisfies the condition $\Delta \phi(j_1,j_2)<$ 2.5. The cut on $\Delta \phi(j_1,j_2)$ is necessary to suppress the QCD multijet background, where large fake $\slashed{E}_T$ may come from inefficient measurement of one of the jets. Furthermore, in order to reduce other backgrounds, such as $W(\rightarrow lv)$ + jets, $Z(\rightarrow l^+ l^-)+$ jets, and $\bar{t}t +$ jets with leptonic top decays, the events containing isolated electrons, muons, taus, or photons with $p_T >$ 20 GeV and $|\eta| < 2.5$ are discarded. 
We then count the events and present the exclusion limits at 95\% C.L. in Fig.~\ref{futurecollider}.

It is obvious from Fig.~\ref{futurecollider} that the sensitivity of collider strongly depends on whether $Z'$ is on shell or not.
When $m_\chi < \frac{1}{2}m_Z' $, $Z'$ is on shell produced and the cross section is resonantly enhanced.
In this case the DM production cross sections and collider sensitivities are almost independent of its mass.
 When  $m_\chi > \frac{1}{2}m_Z' $, $Z'$ is off shell produced, the DM production  cross section  is proportional to $[g_qg_\chi/{(Q^2-m_{Z'}^2)}]^2$ ($Q^2$ is the typical momentum transfer to the DM pair) and is suppressed by $1/Q^2$.
Particularly, for the case $m_{Z'}^2 \ll Q^2$ , the DM cross section is proportional to $[g_qg_\chi/Q^2]^2$  and is irrelevant to $m_{Z'}$, which is demonstrated in the left panel of Fig.~\ref{futurecollider} as that the solid and the dashed lines for the same color appear to close each other with the increase of $m_\chi$.

Compared to direct and indirect detections, the collider search would have stronger capability for the region $m_\chi < \frac{1}{2}m_Z'$. Direct detection will be sensitive for $m_\chi > 10 $ GeV, while indirect detection will be sensitive for $m_\chi > 100$ GeV,   they could probe different mass regions and are complementary to each other.

%%%%%%%%%%%%%%%%%%%%%%%%%%%%%%%%%%%%%%%%%%%%%%%%%%%%
\section{Conclusions}
\label{sec:summary}

%%%%%%%%%%%%%%%%%%%%%%%%%%%%%%%%%%%%%%%%%%%%%%%%%%%%

We constructed a $U(1)'$ model inspired by $E_6$ which has the isospin-violating dark matter.
After a few steps of gauge symmetry breaking, the unbroken gauge symmetry at TeV scale is $SU(3)_C \times SU(2)_L \times U(1)_Y \times U(1)'$. 
For the purpose of phenomenological study, we introduced some new particles  to this model.
Especially, due to the residual $Z_2$ symmetry,  an SM singlet fermion $\chi$ with $U(1)'$ charge ${\bf -27/2}$  is absolutely stable and then a DM candidate.

By choosing a proper linear combination of two extra $U(1)$ gauge symmetries in $E_6$, 
we naturally obtained the ratio $f_n/f_p=-0.7$ so as to maximally relax the constraints 
from the Xenon based  direct detection experiments. 
Compared to isospin-conservation case, the constraints from the Xenon based experiments are 
relaxed by a factor of about $\mathcal{O}(10^4)$.
We studied the sensitivities of dark matter direct and indirect detection experiments, 
and found the parameter spaces that have the observed relic density.
For  $m_\chi \sim \frac{1}{2}m_{Z'}$, the constraints from indirect detection experiments  
are enhanced due to resonance effects.

We then studied the sensitivities of the future colliders with center mass energy $\sqrt{s}$= 33/50/100 TeV.
The sensitivities of the collider searches are highly dependent on whether $Z'$ is on-shell or not.
Moreover, we compared the different detection methods, and showed that
the future colliders will provide the much better searches in our model, 
especially for the region $m_\chi < \frac{1}{2}m_Z'$.

%%%%%%%%%%%%%%%%%%%%%%%%%%%%%%%%%%%%%%%%%%%%%%%%%%%
\begin{acknowledgments}

The research of TL was supported by the Projects 11847612 and 11875062 supported by the 
National Natural Science Foundation of China, and by the Key Research Program of Frontier Science, CAS.
QFX is supported  by the China Postdoctoral Science Foundation under Grant No. 8206300015.
And QSY and XHZ is supported by the Natural Science Foundation of China under the grant No.  11475180 and No. 11875260.

\end{acknowledgments}

%%%%%%%%%%%%%%%%%%%%%%%%%%%%%%%%%%%%%%%%%%%%%%%%%%%%%%%%%%%%%%%%%%%%%%

\appendix
\section{DM annihilation cross sections and relic density}
\label{app:anni}
In our IVDM model, DM annihilates into quarks to realize observed relic density, and the annihilation cross sections are
    	\begin{eqnarray}
	\label{eq:sigma}
    	\sigma_{ann} &=& \sum_q \frac{\beta_q c_q g_\chi^2}{12\pi \beta_\chi ((s-m_{Z'}^2)^2+m_{Z'}^2 \Gamma_{Z'}^2)}((g_{q_V}^2(s+2(m_q^2+m_{\chi}^2)+4\frac{m_q^2 m_{\chi}^2}{s})
    	\nonumber\\&&
    	+g_{q_A}^2(s+4(m_q^2+m_{\chi}^2)+28\frac{m_q^2 m_{\chi}^2}{s}-24\frac{m_q^2 m_{\chi}^2}{m_{Z'}^2}+12\frac{sm_q^2 m_{\chi}^2}{m_{Z'}^4})
    	\nonumber\\&&
    	+2g_{q_V}g_{q_A}(s-(m_q^2+m_{\chi}^2)-8\frac{m_q^2 m_{\chi}^2}{s})),
    	\end{eqnarray}
    	where $s$ is the squared center-of-mass energy of a DM particle pair and color factor $c_q=3$.  $\beta_f=\sqrt{1-\frac{4m_f^2}{m_{Z'}^2}}$ ($f=q$ and $\chi$).
    	
    	The width of $Z'$ can be expressed as  	
    	\begin{eqnarray}
    	\Gamma_{Z'}=\Gamma(Z'\rightarrow \chi\bar{\chi})+\sum_q c_q\Gamma(Z'\rightarrow q\bar{q}),
    	\end{eqnarray}    	
 with
    	\begin{eqnarray}
    	\Gamma(Z'\rightarrow q\bar{q})&=&\frac{m_{Z'}}{12\pi}(g_{q_A}^2\xi_q (1+\frac{2m_q^2}{m_{Z'}^2})+g_{q_V}^2\xi_q^3),\\
	\Gamma(Z'\rightarrow \chi\bar{\chi}) &=&\frac{m_{Z'}}{12\pi}g_{\chi}^2(\xi_\chi (1+\frac{2m_\chi^2}{m_{Z'}^2})+\xi_\chi^3).
    	\end{eqnarray}    	
The particle explanation of $Z'$  is  $\Gamma_{Z'} < m_{Z'}$, which in turn roughly require $g_u < 0.89$.
 	
In order to study DM relic density and indirect detection signals, we need to calculate the thermally averaged annihilation cross section $\langle \sigma_{ann}v_M \rangle$, where $v_M \equiv \frac{\sqrt{(p_1 \cdot p_2)^2-m_1^2m_2^2}}{E_1E_2}$ is the Moller velocity. However, instead of calculating $\langle \sigma_{ann}v_M \rangle$ directly,  it is more convenient to calculate $\langle \sigma_{ann}v_{rel} \rangle$ in the laboratory frame, which means one of the initial particles is at rest, and get the same result. Here $v_{rel}$ is the relative velocity between them.

    	In the laboratory frame, when DM is non-relativistic, $s$ can be expanded as $ 4m_{\chi}^2 + m_{\chi}^2v^2 +\frac{3}{4}m_{\chi}^2v^4 + {\cal O}(v^6)$ , with  $v\equiv v_{rel} =\beta_{\chi}(1-\frac{2m_{\chi}^2}{s})^{-1} $ . 
Plugging  this expression into Eq.~(\ref{eq:sigma}), one can expand $\sigma_{ann}v$ as  $a+bv^2+{\cal O}(v^4)$  with coefficients $a$ and $b$ given by 	
    	\begin{eqnarray}
    	a&=&\sum_q \frac{c_q g_{\chi }^2 \sqrt{1-\frac{m_q^2}{m_{\chi}^2}} (2 g_A g_V (m_{\chi}^2-m_q^2) m_{Z'}^4+g_A^2 m_q^2 (m_{Z'}^2-4 m_{\chi}^2){}^2+g_V^2 (m_q^2+2 m_{\chi}^2) m_{Z'}^4)}{2 \pi  m_{Z'}^4 ((m_{Z'}^2-4 m_{\chi}^2)^2+m_{Z'}^2 \Gamma_{Z'}^2)}\\
	 b&=&\sum_q \frac{v^2 c_q g_{\chi }^2}{48 \pi  m_{\chi }^2 \sqrt{1-\frac{m_q^2}{m_{\chi }^2}} m_{Z'}^4 ((m_{Z'}^2-4{m_{\chi}^2})^2+m_{Z'}^2 \Gamma_{Z'}^2){}^2}
    	\nonumber\\&&
    	 (-2 g_A g_V (m_q^2-m_{\chi }^2) m_{Z'}^4 (m_q^2 (400 m_{\chi }^4+13 m_{Z'}^2 (m_{Z'}^2+\Gamma _{Z'}^2)-152 m_{\chi }^2 m_{Z'}^2)
    	 \nonumber\\&&
    	 +2 m_{\chi }^2 (-80 m_{\chi }^4+m_{Z'}^2 \Gamma _{Z'}^2+16 m_{\chi }^2 m_{Z'}^2+m_{Z'}^4))+g_A^2 (m_q^4 (3840 m_{\chi }^8+16 m_{\chi }^4 (3 m_{Z'}^2 \Gamma _{Z'}^2+98 m_{Z'}^4)
    	 \nonumber\\&&
    	 -8 m_{\chi }^2 (9 m_{Z'}^4 \Gamma _{Z'}^2+38 m_{Z'}^6)+23 m_{Z'}^6 (m_{Z'}^2+\Gamma _{Z'}^2)-3840 m_{\chi }^6 m_{Z'}^2)
    	 \nonumber\\&&
    	 -4 m_q^2 m_{\chi }^2 (768 m_{\chi }^8-4 m_{\chi }^2 (3 m_{Z'}^4 \Gamma _{Z'}^2+20 m_{Z'}^6)+7 m_{Z'}^6 (m_{Z'}^2+\Gamma _{Z'}^2)-768 m_{\chi }^6 m_{Z'}^2+352 m_{\chi }^4 m_{Z'}^4)
    	 \nonumber\\&&
    	 +8 m_{\chi }^4 m_{Z'}^4 (16 m_{\chi }^4+m_{Z'}^2 \Gamma _{Z'}^2-8 m_{\chi }^2 m_{Z'}^2+m_{Z'}^4))
    	 \nonumber\\&&
    	 +g_V^2 m_{Z'}^4 (m_q^4 (368 m_{\chi }^4+11 m_{Z'}^2 (m_{Z'}^2+\Gamma _{Z'}^2)-136 m_{\chi }^2 m_{Z'}^2)
    	 \nonumber\\&&
    	 +2 m_q^2 m_{\chi }^2 (112 m_{\chi }^4+m_{Z'}^2 \Gamma _{Z'}^2-32 m_{\chi }^2 m_{Z'}^2+m_{Z'}^4)
    	 \nonumber\\&&
    	 -4 m_{\chi }^4 (112 m_{\chi }^4+m_{Z'}^2 \Gamma _{Z'}^2-32 m_{\chi }^2 m_{Z'}^2+m_{Z'}^4)))
    	\end{eqnarray}     	

    	To get relic density, we can use an approximate function instead of solving the Boltzmann equation numerically  	
    	\begin{eqnarray}
    	\Omega_{\chi}h^2=2\times 1.04\times 10^9 \mathrm{GeV}^{-1} (\frac{T_0}{2.725~\mathrm{K}})^3 \frac{x_f}{M_{pl}\sqrt{g_*(x_f)}(a+\frac{3b}{x_f})}
    	\end{eqnarray}
    	where $x_f \equiv \frac {m_{\chi}}{T_f} \sim {\cal O}(10)$ , $T_f$ is the DM freeze-out temperature, $T_0=2.725\pm 0.002K$ is the present CMB temperature, and $g_*(x_f)$ is the effective relativistic degrees of freedom at the freeze-out epoch.

%\begin{thebibliography}{99}
%\end{thebibliography}
\bibliographystyle{JHEP.bst}
\bibliography{IVDM}

\providecommand{\href}[2]{#2}\begingroup\raggedright\endgroup

\end{document}